\documentclass[superscriptaddress, dvipsnames, amsmath, amssymb, aps, prl, twocolumn, longbibliography]{revtex4-2}
\usepackage{url}
\usepackage{graphicx}
\usepackage{dcolumn}
\usepackage{bm}
\usepackage{footnote}
\usepackage{mathtools}
\usepackage{psfrag}
\usepackage{amsthm}

\usepackage[colorlinks = true,linkcolor=blue,citecolor=red,urlcolor=magenta]{hyperref}
\usepackage{dsfont}
\usepackage{tabularx}
\usepackage[capitalize]{cleveref}
\usepackage{color}  % for colors
\RequirePackage[dvipsnames,usenames]{xcolor}
\usepackage[T1]{fontenc}
\usepackage{newtxtext}
\usepackage{newtxmath}

\usepackage{booktabs}

\begin{document}

\title{Universal Pareto fronts and trade-off relations in multi-objective \\ optimization of thermal machines}

\author{José A. Almanza-Marrero}
\affiliation{Institute for Cross-Disciplinary Physics and Complex Systems, (IFISC) UIB-CSIC, Campus Universitat de les Illes Balears E-07122, Palma de Mallorca, Spain}

\author{Édgar Roldán}
\affiliation{ICTP – The Abdus Salam International Centre for Theoretical Physics, Strada Costiera 11, 34151 Trieste, Italy}

\author{Gonzalo Manzano}
\email{gonzalo.manzano@ifisc.uib-csic.es}
\affiliation{Institute for Cross-Disciplinary Physics and Complex Systems, (IFISC) UIB-CSIC, Campus Universitat de les Illes Balears E-07122, Palma de Mallorca, Spain}
            
\begin{abstract} 
The performance of thermal machines is characterized with several, often competing quantities, such as power, efficiency, energy waste, and resilience to environmental noise. We employ multi-objective optimization to find the best thermal machines operating in the irreversible linear-response regime. We derive exact analytical parameterizations for the optimal (Pareto) fronts associated with mean extracted power, efficiency, entropy production and the amplitude of power fluctuations. {These optimal fronts set the quantitative fundamental limits for the performance of linear-response machines, which are achieved under tight coupling conditions (endoreversible machines). The geometry of the fronts is universal:} two-, three-, and four-objective trade-offs follow analytical formulae that do not depend on the value of any physical parameter of the machine. 
{Furthermore, we test our results with existing experimental data from different physical systems operating also beyond the linear regime, ranging from atomic to macroscopic scales, and including single-atom engines, colloidal systems, macroscopic engines and  power plants.}
\end{abstract}

\maketitle

Thermal machines are physical systems that, when fueled by input energy,  perform  output tasks such as heat pumping or the production of work. 
Carnot's pioneering work in the 18th century discovered fundamental constraints in the performance of such machines
that cannot be overcome~\cite{Carnot1824}.
Consider, for instance, a steam engine operating between a hot source and a cold sink: it can only approach its maximum possible efficiency—the Carnot efficiency—if it runs extremely slowly, almost quasistatically, yielding zero power in the ideal limit. Pushing the machine to run faster increases its power output, but also inevitably increases the amount of energy dissipated~\cite{curzon75}.
From Carnot’s discoveries to contemporary analyses of nanomachines, the relationship between maximum efficiency and maximum power regimes has been studied extensively~\cite{Novikov58,Kosloff84,Leff87,broeck05,schmiedl08,esposito2009,Allahverdyan13,Proesman16,korbel2026quo}. In parallel, technological advances over the past few decades have brought access to  manipulation of natural and artificial  microscopic machines with exquisite accuracy, ranging from single atoms~\cite{Rossnagel16,Bouton21} to molecular motors~\cite{kolomeisky2007molecular,seifert12,blickle12,Martnez16}. These small machines also obey power–efficiency and power–dissipation trade-offs like macroscopic machines, but with an additional constraint: they must operate reliably in the presence of fluctuations of different origins (thermal, chemical, etc.) present in their surrounding environment~\cite{jarzynski97,crooks99,bustamante2005nonequilibrium,barato15,gingrich16}.

\begin{figure} 
  \centering
  \includegraphics[width=1.0\linewidth]{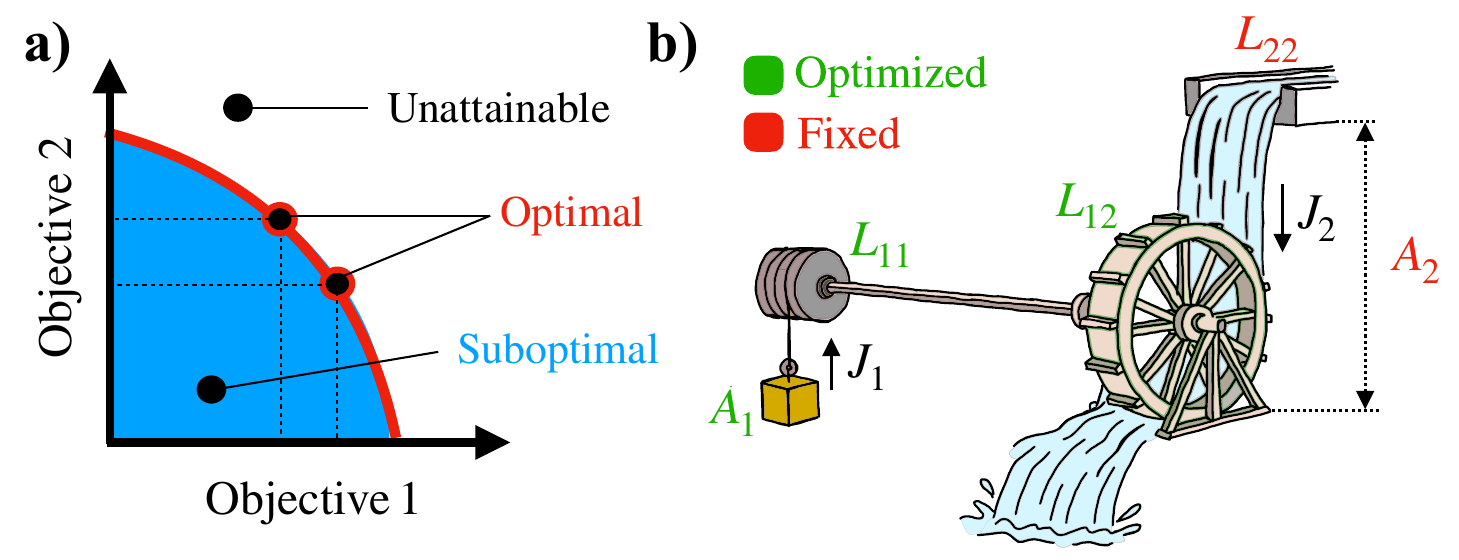}
  \caption{(a) Multi-objective optimization for the case of two objectives.
  The blue region represents the permitted values for the two objectives while the white region represents unattainable configurations. The red solid line corresponds to  optimal configurations (\textit{Pareto} front). 
  (b) Sketch of a linear-response watermill. Parameters $A_1$ (the weight of the load), $L_{11}$ (the efficacy of the pulley mechanism) and $L_{12}$ (the response of the  watermill blades) can be typically engineered. Parameters $A_2$ (the height of the waterfall) and $L_{22}$ (hydraulic conductivity of input water flow) represent thermodynamic resources available to the watermill. Under tight coupling conditions (endoreversible machines) $L_{11}$ and $L_{12}$ become proportional to each other for fixed $L_{22}$.} \label{fig: 1}
\end{figure}

 A natural way to tackle competing objectives is through the so-called multi-objective (so-called \textit{Pareto}) optimization. This idea originally emerged in economics~\cite{koopmans51}, where one seeks to balance different goals—such as cost and quality—by assuming that any desired output can be maximized while compromising at least one of the others through trade-off relations. Although Vilfredo Pareto's socio-economic study~\cite{pareto1935mind} did not explore or venture trade-off relations to be applicable in the natural sciences, Pareto optimization has emerged as a powerful technique, finding widespread applications in miscellaneous disciplines such as engineering, biology, computer science, and  evolutionary theory~\cite{deb01,seoane15}. In thermodynamics, optimization approaches are increasingly being used in a wide range of contexts, including inference, enhancement of free-energy harvesting, and optimal control at the microscale~\cite{Miangolarra25,Pinero24,Loos24}. In this context, multi-objective optimization offers a powerful framework to characterize trade-offs between thermodynamic quantities that typically compete with one another~\cite{yilmaz06,detomas12,ashida21,solon18,erdman23,Foro25,Izumida24}. The main idea is rather intuitive: instead of searching for a single “best” machine, one identifies those designs that cannot be improved in one performance metric—say, power—without compromising another performance metric, such as efficiency, dissipation, or precision. When all such possibilities are considered, they form what is known as a Pareto front. From this perspective, thermodynamic trade-offs provide the appropriate framework to establish the boundaries of machines' performance limits (see Fig. \ref{fig: 1}a).

In this work, we derive the exact optimal trade-off relations between power, efficiency, dissipation, and precision for generic thermal machines operating in the linear irreversible response regime~\cite{prigogine67,degroot84}. We find that for endoreversible machines (the class originally considered by Carnot) these optimal relations are universal, and take the form of rather compact analytical expressions. { Importantly, our results also show that these expressions  establish upper bounds for all other machines in the linear regime, hence setting their ultimate achievable thermodynamic limits.}
Beyond the familiar results of maximum power and Carnot efficiency, { our results comprise the derivation of several other tradeoff relations, which we show to be independent of the specific physical details of the machine. This includes a new derivation of the thermodynamic uncertainty relation (TUR)~\cite{barato15,gingrich16,pietzonka18}, together with the conditions for their saturation in the linear regime, resulting from the multi-objective optimization framework. All in all, our results provide a unifying approach for assessing the performance of thermal machines operating in the linear response regime.} 
Together with the mathematical simplicity of our results, {  the universality of our bounds} greatly enhances their potential applications. As a matter of fact, we {  test} 
our theoretical results using experimental data existing in the literature from different platforms across scales, ranging from single-atom engines to power plants, operating even beyond the linear regime.

 {\it Linear irreversible thermodynamics}.--
 Thermal machines often operate by exploiting spatial or temporal inhomogeneities of physical quantities such as  temperature, chemical potential, or pressure. Such gradients induce thermodynamic forces, denoted here by $A$, which drive the machine out of equilibrium. Each force is associated with a conjugate flux~$J$, such as heat or particle currents. The main goal of  a thermal machine is to convert one type of energy (e.g. heat) into another type (e.g. work). In the simplest scenario, this conversion requires  at least  two forces: a load (input) force $A_{1}$ against which useful work is performed, and a driving (output) force $A_{2}$ which supplies energy to the system. The working regime is characterized by the sign of the fluxes: an entropy-producing flux in the direction of the driving force, $A_2 J_2 \geq 0$, sustains the  other flux against the load force, $A_1 J_1 \leq 0$ (see the End Matter for prototypical examples).
 
 In this context, one of the key quantities of interest is the average power extracted from the machine $P$, that is, the work that the load force is able to produce per unit time~\cite{broeck05}
\begin{equation}\label{eq:power}
    P \equiv -T~ A_1 J_1, 
\end{equation}
where $T$ is a reference temperature of the power-producing device~\cite{manzano20}. Another   key quantity characterizing a thermodynamic process is the dissipation rate ${\Sigma}$ (or entropy production rate), which in the linear-response regime can be expressed as a combination of fluxes and forces~\cite{prigogine67,degroot84}
\begin{equation}\label{eq:dissipation}
    {\Sigma} = A_1 J_1 + A_2 J_2 \geq 0,
\end{equation}
where the last inequality enforces the second law of thermodynamics. From the second law, it is also natural to introduce a measure of (normalized) efficiency $\bar{\eta}$ of the machine  
\begin{equation}
\overline{\eta} \equiv -\dfrac{A_1 J_1}{A_2 J_2}, 
\end{equation}
which captures the fraction of entropy flux that is used to generate power. 
In the working regime, $P\geq 0$ and $0 \leq \overline\eta \leq 1$. The maximum efficiency $\overline\eta = 1$, is reached in the reversible limit corresponding to zero dissipation $\Sigma = 0$, which leads to Carnot's bound for heat engines~\footnote{For heat engines, the reversible limit is associated with the Carnot limit, since in  that case $J_2 = Q_{\rm h}$ is the input heat from a hot source at $T_{\rm h}$ and $A_2 = 1/(T_{\rm h} - T)$ is the inverse temperature difference, leading to the energetic efficiency 
$\eta_{\rm e} \equiv \frac{P}{Q_{\rm h}} = \overline\eta~\left(1 - \frac{T}{T_h}\right) \leq  \eta_C$,
which reaches Carnot's bound $\eta_C= 1 - (T/T_h)$ for $\overline\eta = 1$.}.

When forces are small, fluxes respond linearly, and Onsager theory holds, $\vec{J}=\mathbb{L} \vec{A}$,~\cite{onsager31a}
\begin{equation}
    \begin{pmatrix}
J_1 \\
J_2
\end{pmatrix}
=
\begin{pmatrix}
L_{11} & L_{12} \\
L_{21} & L_{22}
\end{pmatrix}
\begin{pmatrix}
A_1 \\
A_2
\end{pmatrix}, 
\end{equation}
where the matrix elements $L_{ij}$ in $\mathbb{L} $ are known as  Onsager coefficients. The second law of thermodynamics [Eq.~\eqref{eq:dissipation}] constrains  $\mathbb{L}$ to be positive semi-definite, which implies: $L_{11}\geq 0$, $L_{22}\geq 0$, and $(L_{11}+L_{22})^2\leq 4 L_{12}L_{21}$. Under time-reversal symmetry conditions (e.g. in absence of magnetic fields), $\mathbb{L}$ is symmetric, leading to the celebrated Onsager reciprocity $L_{12} = L_{21}$. If additionally one has $L_{12}^2=L_{11}L_{22}$, the two currents become proportional, enabling the machine to approach reversibility (Carnot's bound) at vanishing dissipation~\cite{VdenBroeck2007}. The machines verifying this additional property are called endoreversible machines~\cite{callen85}{ , but the same condition has been also often refereed to as tight (or strong) coupling~\cite{Parmeggiani1999,Parrondo2002,broeck05}}.

On mesoscopic scales, fluctuations in currents play a fundamental role, leading to a noisy output work signal characterized by its variance, $\sigma^2_P \equiv \langle P^2\rangle - \langle P \rangle^2$. Keeping a small output-power variance is key to ensure resilience to environmental fluctuations, which may otherwise spoil the functionality of the machine. Within the linear response regime, the fluctuation–dissipation theorem~\cite{callen85} relates variances of currents to Onsager coefficients, implying that the equilibrium variance of each current satisfies $\langle J_i^2\rangle_{\mathrm{eq}} = 2 L_{ii}$. Under the standard assumption that forces are externally imposed and effectively nonfluctuating (differences in temperatures, chemical potentials, etc), the output-power variance obeys
\begin{equation}\label{eq:5}
    \sigma^2_P =
    A_1^{2}T^{2} \sigma^2_{ J_1} = 2 T^{2}A_1^{2} L_{11},
\end{equation}
where $\sigma^2_{J_{1}} = \langle J_1^2 \rangle_{\mathrm{eq}} -\langle J_1\rangle_{\mathrm{eq}} ^2 $ and $\langle J_1\rangle_{\mathrm{eq}} = 0$. Notably, Eq.~\eqref{eq:5} allows treating $\sigma^2_P$ on the same footing as the other performance metrics $P$, $\eta$, and ${\Sigma}$ in a multi-objective optimization analysis.

\begin{figure}[t]
\centering\includegraphics[width=0.95 \linewidth]{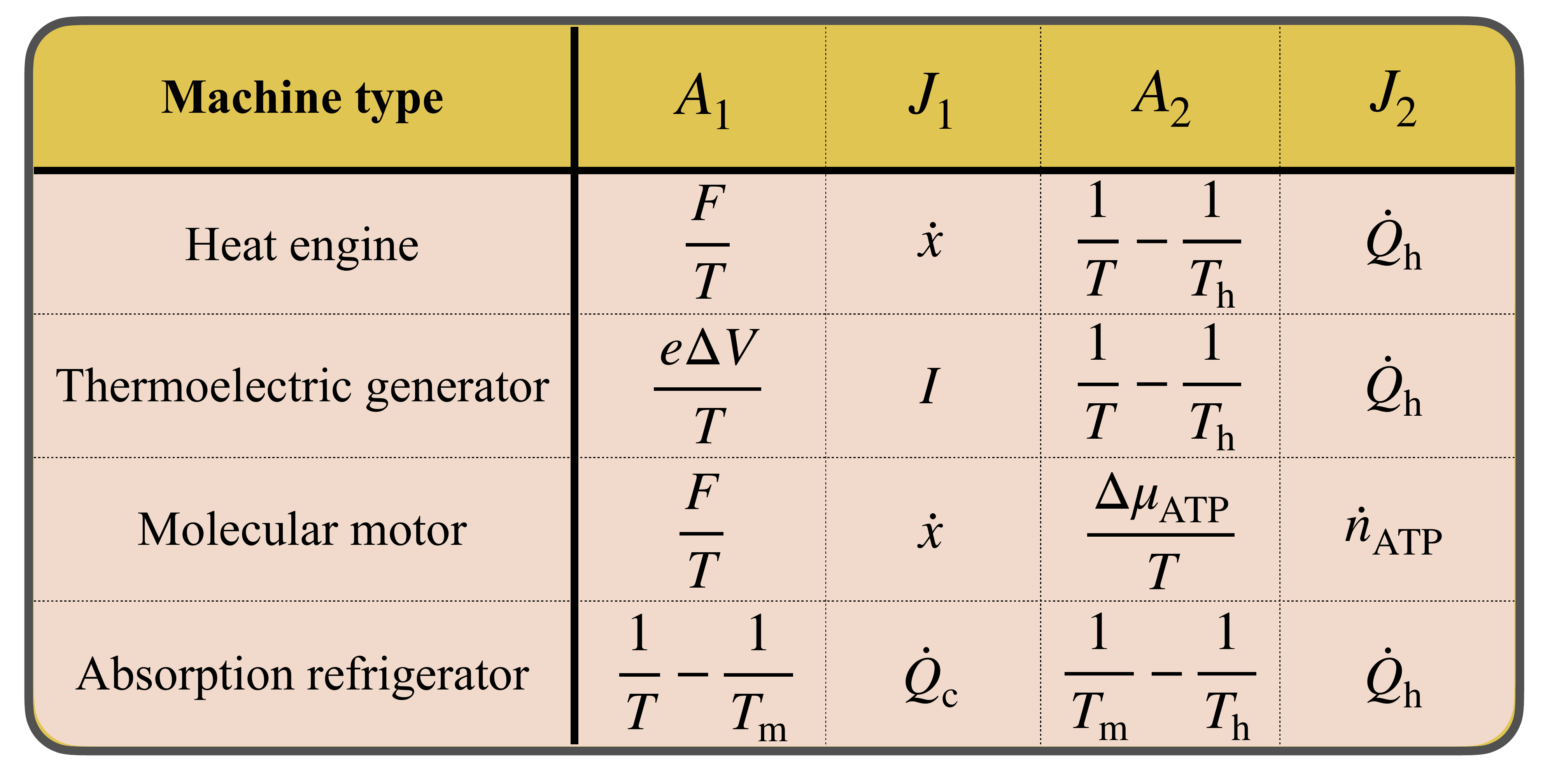}
\caption{Onsager affinities $A_1$, $A_2$ and their conjugate fluxes $J_1$, $J_2$ for different thermal machines. Here $F$ is an external force, $\dot{x}$ the linear velocity, $e$ the elementary charge, $\Delta V$ the electrical potential difference, $I$ the electric current, $\Delta \mu_{\mathrm{ATP}}$ the chemical potential difference per ATP, and $\dot{n}_\mathrm{ATP}$ the ATP consumption rate. $T_{\mathrm{h}}$, and $T_{\mathrm{m}}$ denote hot and intermediate reservoir temperatures, while $\dot{Q}_{\mathrm{h}}$ and $\dot{Q}_{\mathrm{c}}$ are heat fluxes from the respective reservoirs.}
  \label{fig:table}
\end{figure}

An insighful picture to illustrate heat engines are water mills driven by a waterfall, where the falling water represents the heat flowing from one reservoir to another (see Fig.~\ref{fig: 1}b). 
Within linear thermodynamics, the affinity that allows work to be extracted corresponds to the height of the waterfall ($A_2$) and its associated flux ($J_2$) is represented by the stream of water flowing down the cascade. These resources set the mill into motion and allow it to perform a useful task, such as lifting a mass load. The weight of this load represents the second affinity ($A_1$), while its corresponding flux, $J_1$, is the speed at which the load is lifted. 
In this picture, $L_{22}$ is related to the water losses (hydraulic conductivity) of the input flow atop the waterfall, while $L_{12}$ quantifies how efficiently the mill blades can convert falling water into motion. 
On the other hand, $L_{11}$ measures the efficacy of the pulley mechanism, determining how easily the mill motion is used to lift the load. {As we will shortly see, $L_{22}$ and $A_2$} represent the {\em resources} available to the machine. This is rather intuitive {  in the watermill picture:} we cannot choose the river that feeds the waterfall or its height, but we are free to design the mill itself. {  Additionally, in Fig.~\ref{fig:table} we provide a  short list of prototypical examples of thermal machines operating in the linear response regime. We show the input $(J_2)$ and output $(J_1)$ fluxes, together with their corresponding affinities $A_1$ and $A_2$ for a mechanical heat engine, a thermoelectric generator, a Kinesin molecular motor, and an autonomous absorption refrigerator.}

{\it   Optimizing linear-response thermal machines}.--
In order to address the joint optimization of efficiency $\eta$, dissipation $\Sigma$, mean power $P$, and power fluctuations $\sigma_P^2$, we employ a weighted-sum scalarization, where a parametrization with three independent  weights $0\leq \alpha\leq 1$, $0\leq \beta\leq 1$, and $0\leq \gamma\leq 1$ is introduced, together with a fourth weight \(1-\alpha-\beta-\gamma\) that ensures normalization. We then proceed to the optimization of a single objective function of the form 
\begin{equation}
\Omega(\alpha,\beta,\gamma) = \alpha~\overline{P} + \beta~\overline{\eta} - \gamma~\overline{\Sigma} - (1-\alpha-\beta-\gamma)\overline{\sigma^2_{P}},  \label{eq:omega}
\end{equation}
where the minus sign stands for the thermodynamic quantities that have to be minimized and we have $\overline{P} = P/P_{\rm max}$, $\overline{\eta} = \eta/\eta_{\rm max}$,~$\overline{\Sigma} = \Sigma/\Sigma_{\rm max}$ and $\overline{\sigma^2_{P}} = \sigma^2_P / \sigma^2_{P,{\rm max}}$ being the  thermodynamic quantities normalized by their respective global maxima. Maximizing~$\Omega$ in Eq.~\eqref{eq:omega} is a weighted  joint optimization problem towards maximum power, maximum efficiency, minimum dissipation and minimum power fluctuations{  , which requires scanning the relative weights associated with every specific thermodynamic quantity  ---$\alpha$ for the mean power, $\beta$ for the efficiency and $\gamma$ for the dissipation--- in the simplex $\alpha+\beta+\gamma\leq 1$.}

{  Although the optimization problem in Eq.~\eqref{eq:omega} in not analytically solvable in general, a crucial technique is to  identify endoreversible machines, that is, those verifying both Onsager reciprocity $L_{12} = L_{21}$ and the tight coupling condition $L_{12}= \sqrt{L_{11}L_{22}}$, as those providing, at the same time, maximum power, maximum efficiency, and minimal dissipation for a fixed value of power fluctuations. In the Supplementary Material, we prove that any reciprocal thermal machine~\footnote{We refer to reciprocal thermal machines as those machines that satisfy the Onsager reciprocity condition $L_{12} = L_{21}$.} in the linear-response must satisfy the following bounds:
\begin{equation}
P\left(q\right) \le P_{\mathrm{endo}}, ~~
\eta(q) \le \eta_{\mathrm{endo}}, ~~
\Sigma(q) \ge \Sigma_{\mathrm{endo}},
\label{eq:bounds_envelopes}
\end{equation}
and $\sigma^2_P(q) = \sigma^2_{P~ {\rm endo}}$. Above $q = |L_{12}|/\sqrt{L_{11}L_{22}}$ is a dimensionless parameter measuring the coupling between the fluxes and we used the label ``endo'' to denote the corresponding values for endoreversible operation. Inequalities~\eqref{eq:bounds_envelopes} saturate  only under perfect (tight) coupling (\(q=1\)) or in the trivial zero-power limit. Notably, we find that these bounds remain valid even when Onsager reciprocity is broken, provided the asymmetric coefficients of the Onsager matrix \(\mathbb{L}\) satisfy \(|L_{12}|\leq \sqrt{L_{11}L_{22}}\) and ,\(|L_{21}|\leq \sqrt{L_{11}L_{22}}\).}

\begin{figure*}[t]
  \centering\includegraphics[width=0.92\textwidth]{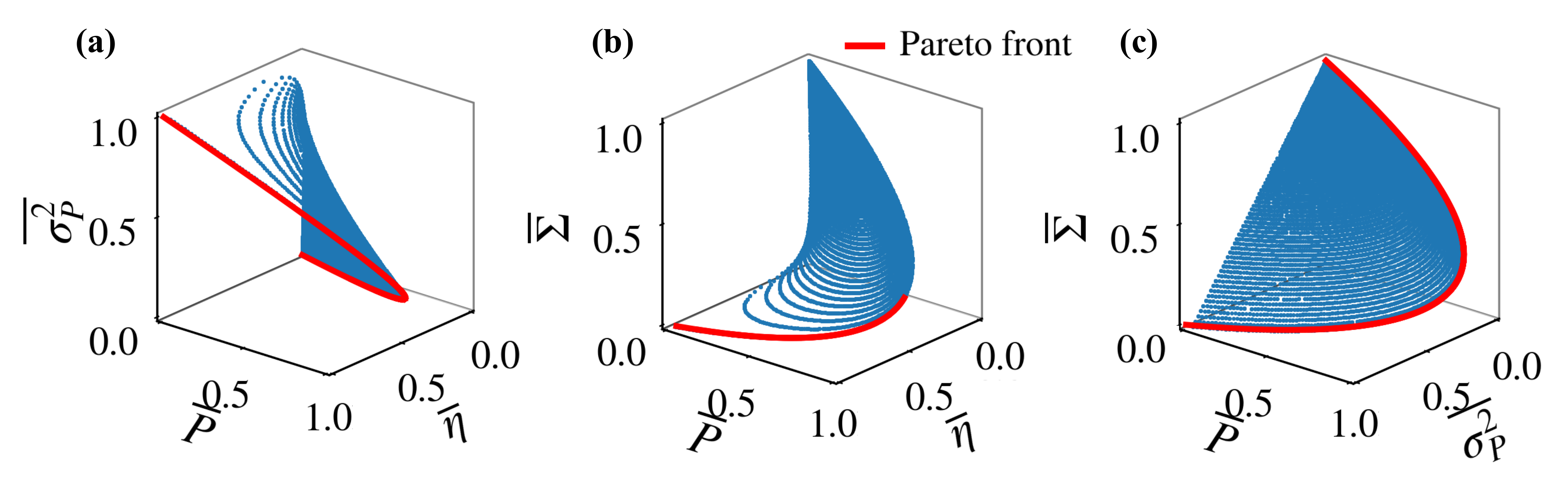}
  \caption{(a-c) Pareto fronts (red solid lines) in three dimensional spaces derived in Eq.~\eqref{eq:tur}. The blue points are  configurations sampled within the linear regime using fixed parameters $A_2=1$, $L_{22}=1$.
  }
  \label{fig: 2}
\end{figure*}

{  Using Eqs.~\eqref{eq:bounds_envelopes} we can now focus on endoreversible machines, which allow us to obtain an analytical solution of the multi-objective optimization problem}. For these machines, the maximal values of mean power, dissipation, and power fluctuation depend only on $A_2, L_{22}$ and $T$; they read $P_{\rm max} \equiv \max(P) = T A_2^2 L_{22}/4$, $\Sigma_{\rm max} \equiv \max(\Sigma) = A_2^2 L_{22}$, $\sigma_{P, {\rm max}}^2\equiv \max(\sigma_{P}^{2}) = 2 T^{2} A_2^2 L_{22}${ , and $\eta_{\rm max}=1$}. We solve this multi-objective optimization problem analytically for all values of $\alpha,\beta$ and $\gamma$ in the simplex by maximizing $\Omega$ with respect to $A_{1}$ and $L_{11}$ {  (the design variables) while fixing $L_{22}$ and $A_{2}$  
(the resources). This choice is justified by the fact that all (maximum) performance metrics in the endoreversible limit scale with $L_{22}$ and $A_{2}$, and varying them would make the problem in Eq.~\eqref{eq:omega} untractable. Nevertheless, as shown in the SM, our result in Eq.~\eqref{eq:bounds_envelopes} provides an alternative route to establishing Pareto optimality in this case.}

{Maximizing Eq. \eqref{eq:omega} over the physical operating regime with $\alpha,\beta,\gamma\geq 0$ and $\alpha + \beta+\gamma \leq 1$, gives the following piecewise solutions: 
\begin{eqnarray}
        \overline{P}\,^{\star}(\alpha,\beta,\gamma) &=& 
    \begin{cases}
        \displaystyle
        \frac{(4\alpha+\beta+2\gamma)
              k}
             {(3\alpha-\beta+1)^2}
        & \beta<\beta_{\rm r} \\[8pt]
        0,
        & \beta\geq\beta_{\rm r}
    \end{cases}
,\label{eq:9}\\
        \overline{\eta}\,^{\star}(\alpha,\beta,\gamma) &=& \begin{cases}
        \displaystyle
        \frac{4\alpha+\beta+2\gamma}
             {6\alpha-2\beta+2}
        & ~~~~ \beta<\beta_{\rm r} \\[8pt]
        1,
        & ~~~~ \beta\geq\beta_{\rm r}
    \end{cases} ,\\
        \overline{\Sigma}\,^{\star}(\alpha,\beta,\gamma) &=& \begin{cases}
        \displaystyle
        \frac{(2\alpha-3\beta-2\gamma+2)^2}
             {4(3\alpha-\beta+1)^2}
        & \beta<\beta_{\rm r} \\[8pt]
        0,
        & \beta\geq\beta_{\rm r}
    \end{cases},\\
        \overline{\sigma^2_P}\,^{\star}(\alpha,\beta,\gamma)&=& \begin{cases}
        \displaystyle
        \frac{(4\alpha+\beta+2\gamma)^2}
             {(6\alpha-2\beta+2)^2}
        & \beta<\beta_{\rm r} \\[8pt]
        1
        & \beta\geq\beta_{\rm r}
    \end{cases},\label{eq:12}
    \end{eqnarray}
with $k=(2\alpha-3\beta-2\gamma+2)$. Here above $\beta_{\rm r} \equiv 2(\alpha - \gamma + 1)/3$ corresponds to the value of the weight $\beta$ from which the reversible (Carnot) point ($\overline{\eta}=1$) becomes the absolute optimum, even if the other weights are not set to zero. The corresponding optimal relation between the load $A_1$ and the parameter $L_{11}$ at the Pareto front is:
\begin{equation}
 A_1^{\star} = A_2\sqrt{\dfrac{L_{22}}{L_{11}^\star}} 
        \left(\dfrac{4\alpha + \beta + 2\gamma}{6\alpha-2\beta +2}\right),
\end{equation}
for $\beta\leq \beta_{\rm r}$, and $A_1^{\star} = A_2\sqrt{L_{22}/L_{11}^\star}$ for the reversible point, which directly tell us how the optimal machines have to be constructed for every choice of the weights.

Equations (\ref{eq:9})--(\ref{eq:12}) describe the optimal performance given by a four-dimensional Pareto front of power, efficiency, dissipation and power fluctuations (see SM for a detailed derivation).  
Interestingly, the shape of the optimal solutions does not depend on any physical detail of our system, but  only dictated by a non-linear combination of the optimization weights. In other words, within the linear regime and under tight coupling (endoreversible operation), the optimal multi-objective compromises between the main thermodynamic performance quantifiers of the machine are described by a universal, model-independent relation. }

{\it   Bounds and tradeoffs for generic thermal machines}.--
In an endoreversible heat engine, heat  and work-producing fluxes are perfectly coupled, i.e.  all the input heat  is converted  into useful work. However, in real systems, internal irreversibilities weaken this coupling. For example, in thermoelectric generators where the electrical current (work flux) and heat flow are ideally linked through the Seebeck effect, parasitic heat conduction may arise, allowing heat to flow without generating electrical power. Likewise, in mechanical heat engines, friction, turbulence, finite-rate heat transfer, or fluid leakage cause part of the energy flow to dissipate as waste rather than being converted into useful work. These leakages reduce the degree of coupling between fluxes and therefore drive the system away from the endoreversible limit. 

\begin{figure*}[t]
\centering\includegraphics[width=1\textwidth]{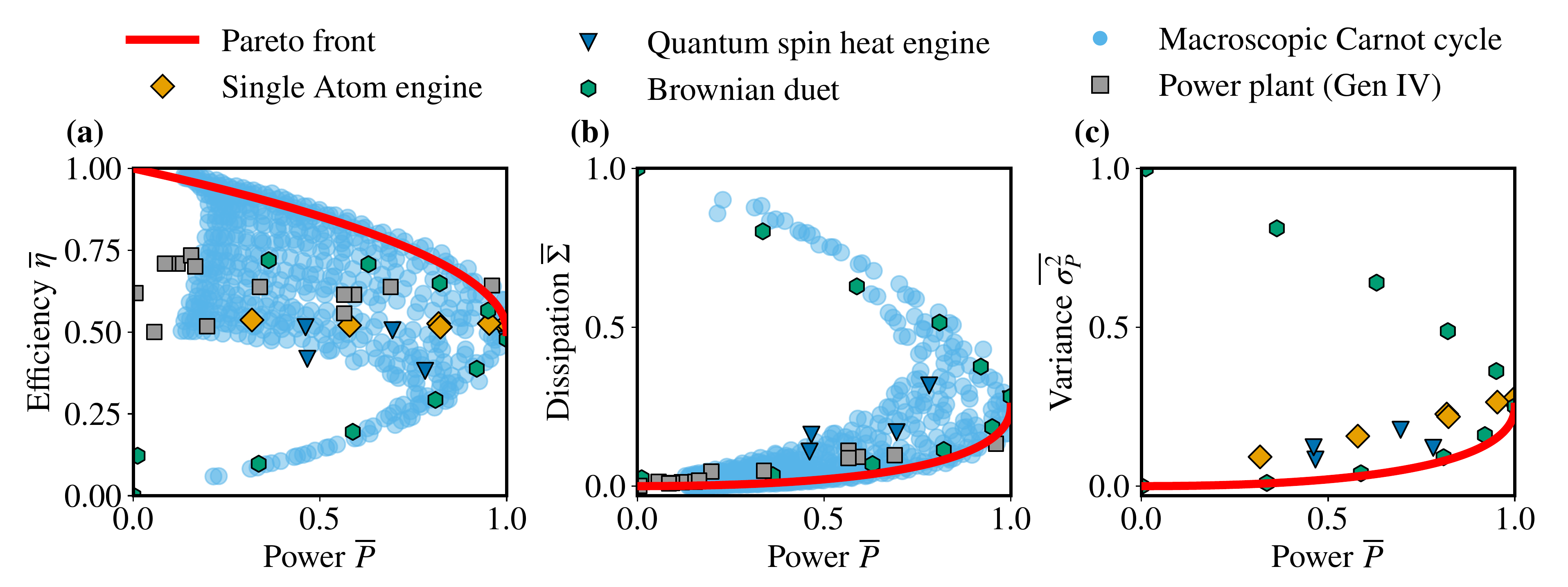}
  \caption{(a–c) Comparison of theoretical Pareto fronts (red solid lines) for the two-dimensional trade-offs derived in Eq.~\eqref{eq:7} with experimental data (colored symbols) from different systems. As indicated in the legend, the data correspond to a single-atom engine~\cite{Bouton21}, a quantum spin heat engine~\cite{Peterson19}, a Brownian engine~\cite{Proesmans16}, an implementation of a finite-time Carnot cycle~\cite{Zhai23} and power plants~\cite{Cui25}.  
  }
  \label{fig: experiemntal_data}
\end{figure*}

An alternative insightful perspective {  of the multiobjective optimization problem} is to write the solutions in Eqs.~(\ref{eq:9})--(\ref{eq:12}) explicitly and interpret them as inequalities {  providing bounds delimiting the ultimate performance limits for the broad class of thermal machines in the linear response satisfying inequalities~\eqref{eq:bounds_envelopes}. The equality is reached under} optimal operation, i.e. over the Pareto front (see Figs.~\ref{fig: 2} and \ref{fig: experiemntal_data}). Such inequalities are not mere mathematical statements but quantitative expressions of physical limits~\cite{jarzynski2012equalities,seifert2019stochastic}. Projecting the multi-objective parametric solution [Eqs.~(\ref{eq:9})--(\ref{eq:12})] onto two-dimensional subspaces { (by setting to zero the corresponding weights)} yields: 
\begin{equation}\label{eq:7}
\begin{split}
&\overline{\eta} \leq \dfrac{1+\sqrt{1-\overline{P}}}{2}, \quad \overline{\Sigma} \geq \dfrac{\big(1-\sqrt{1-\overline{P}}\big)^2}{4},\; \\ & \qquad \qquad \overline{\sigma_{P}^{2} }\geq \dfrac{\big(1-\sqrt{1-\overline{P}}\big)^2}{4}.
\end{split}
\end{equation}
The right-hand side of each inequality  represents the best possible performance, acting as either an upper bound (if we are maximizing the quantity) or a lower bound (if we are minimizing the quantity). 
For example, the first inequality in~\eqref{eq:7}  sets the efficiency at (unconstrained) maximum power ($\overline{P}=1$) to be universally bounded by half of the maximum efficiency, $\overline{\eta}_{MP} \leq 1/2$, as first noticed in Ref.~\cite{broeck05}, and predicts a square-root dependency  of the maximal efficiency as a function of the 'sacrificed' power $1-\overline{P}$. Such result was  derived in Ref.~\cite{Holubec16} for the special case of heat engines. On the other hand, the second and third inequalities in~\eqref{eq:7}  set the minimum  values $\overline{\Sigma}_{MP} \geq 1/4$ and $\overline{\sigma^2_P}_{,MP} \geq 1/4$ for dissipation and  power fluctuations at maximum power, which may be reduced through a non-linear relation  with the amount of sacrificed   output power $1-\overline{P}$. Among other potential implications, we  observe that no output power is possible $\overline{P} = 0$ at maximum efficiency ($\overline{\eta} = 1$), zero dissipation ($\overline{\Sigma} = 0$) or zero power fluctuations ($\overline{\sigma_P^2} = 0$). Indeed, negligible fluctuations may  only be achieved at zero  power. Inequalities~\eqref{eq:7} generalize previous insights on maximum power, maximum efficiency, and minimum dissipation derived in the linear regime in Refs.~\cite{broeck05,Proesman16,Proesmans16,Ma18} by both including fluctuations and providing a whole family of optimal trade-offs between the performance metrics of interest. 

Our analytical formulae permits also tackling the more challenging scenario of the three-dimensional trade-offs shown in Fig. \ref{fig: 2}. In this case, {  projecting the four-dimensional fronts into three-dimensional spaces} we obtain the universal tradeoff relations
\begin{equation}\label{eq:tur}
\begin{split}
T\dfrac{P}{\sigma^2_{P}} \left( \dfrac{\eta}{\eta_{\rm max}-\eta} \right) \leq \dfrac{1}{2}\, , \qquad
&\Sigma~\dfrac{\sigma^2_{P}}{P^2} \geq 2\, , \\ 
\Sigma~\geq \dfrac{P^{2}}{4 T P_{\rm{max}}} \left( \dfrac{\eta_{\rm{max}}}{\eta} \right)^2,
\end{split}
\end{equation}
where we used the relations between the maximal thermodynamic quantities for endoreversible heat engines, $\Sigma_{\rm max} = 4 P_{\rm max} /T$ and $\sigma_{P, {\rm max}}^{2} = 8T P_{\rm max}$, to recast the bounds in terms of unnormalized quantities. The first and second inequalities in Eq.~\eqref{eq:tur} correspond to two analogous formulations of the so-called thermodynamic uncertainty relation (TUR)~\footnote{Notably, if we try to obtain the explicit expression for the full optimization problem [Eqs.~\eqref{eq:9}--\eqref{eq:12}], the second inequality in Eq.~\eqref{eq:tur}[$\Sigma \sigma^{2}_P/P^{2}\geq 2$] is recovered again.}, discussed respectively in Refs.~\cite{pietzonka18} and~\cite{barato15,gingrich16}. The TUR implies that improving the accuracy of the output power (i.e. lowering the fluctuations for a given output power) can only be achieved at the cost of  increasing the dissipation or sacrificing efficiency. On the other hand, the third inequality in~\eqref{eq:tur}  
{  provides a lower bound on the machine dissipation in terms of a quadratic function of its output power over the efficiency. Under tight coupling conditions, where the equality $\overline{\Sigma}=(1-\overline{\eta})^2$ is verified, that relation can be reduced to the power-efficiency tradeoff in Eq.~\eqref{eq:7}}.
By definition, inequalities~\eqref{eq:tur} are saturated only by the optimal {  endoreversible} configurations  (i.e, those on the Pareto front), thus saturating the TUR is optimal in the linear response limit. Moreover, the parametric solutions [Eqs.(\ref{eq:9})--(\ref{eq:12}] obtained from the different optimization schemes identify the specific ways of approaching the optimal limit, depending on how power, efficiency, dissipation, and fluctuations are weighted.

{\it Testing the performance of real-world thermal machines.---} The analytical simplicity of our results, combined with their independence from system-specific details, enables their application to a broad range of real-world machines. In this section, we put our findings to the test across scales, by inspecting published experimental data---from microscopic to macroscopic systems---and across operational regimes ranging from low dissipation to strongly nonequilibrium conditions.  Figure~\ref{fig: experiemntal_data} displays the optimal {  linear-response} trade-offs predicted by Eq.~\eqref{eq:7} as a red solid line. The symbols are obtained through a thorough survey of existing experimental data from different platforms: a single-atom engine composed of a cesium impurity embedded in an ultra-cold rubidium gas~\cite{Bouton21}; an Otto cycle implemented in a liquid-state Nuclear Magnetic Resonance platform using the nuclear spin of a carbon-13 nucleus in a chloroform molecule~\cite{Peterson19}; a Brownian {  duet} heat engine using a single colloidal particle trapped by optical tweezers in a fluid and driven by two phase-shifted periodic forces~\cite{Proesmans16}; an experimental implementation of a macroscopic finite-time Carnot cycle using dry air inside a piston~\cite{Zhai23}; and recently reported performance data from several generations of nuclear power plants~\cite{Cui25}. 

{  We emphasize that the purpose of this test is mostly exploratory, since the data of the real-world machines reported in Fig.~\ref{fig: experiemntal_data} does not necessarily correspond to the linear-response regime where our analytical results are obtained. Surprisingly however, we find that the observed performance of these machines remains always within the predicted bounds in Eqs.~\eqref{eq:7} and \eqref{eq:tur}.} Across the three performance trade-offs shown in Fig. \ref{fig: experiemntal_data}(a-c), we find a consistent hierarchy in how closely different engines approach optimal operation. Finite-time Carnot cycle implementation uniquely span the full optimal-performance branches in the power-efficiency and power-dissipation cases, while the Brownian {  duet} engine and the last generation of nuclear power plants systematically operates near all this corresponding optimal frontiers. Atomic engines display intermediate performance in the power–efficiency figure, and the quantum spin engine remains comparatively further from optimality in the power–dissipation trade-off. At mesoscopic scales where fluctuations are relevant, the Brownian {  duet} engine nearly saturates the optimal bound, with quantum engine implementations exhibiting similarly fluctuating behavior close to the optimal Pareto fronts.

\begin{figure}[t]
\centering\includegraphics[width=0.65\linewidth]{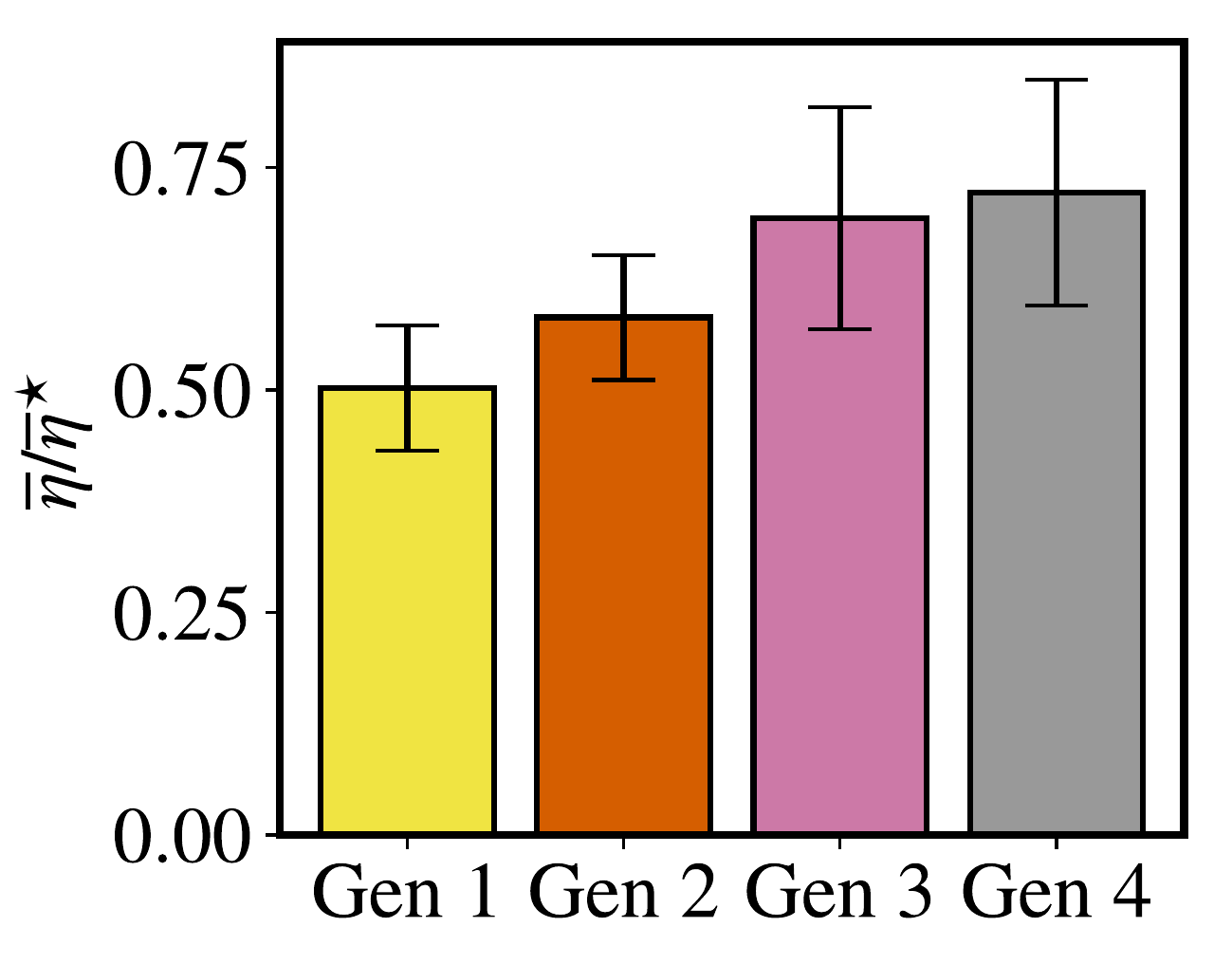}
  \caption{Bar plot of the ratio of the achieved efficiency to the optimal efficiency for different generations of nuclear power plants, based on data reported in Ref.~\cite{Cui25}. For each generation, values are obtained by averaging over individual plants, with error bars indicating the corresponding standard deviations.
  }
  \label{fig: experiemntal_data_2}
\end{figure}

{  Figure~\ref{fig: experiemntal_data_2} illustrates the average distance to optimal performance of different generations of nuclear power plants, using data from Ref.~\cite{Cui25}: Generation I, developed from the 1950s to the late 1960s, introduced the first experimental and commercial reactors; Generation II, from the late 1960s to the 1990s, incorporated larger power outputs with standardized designs; Generation III, from the 1990s onward, added better fuel efficiency and longer operating lifetimes; and Generation IV introduced higher operating temperatures, advanced coolants and improved fuel utilization. For each plant, we compute the fraction of the optimal efficiency achieved $(\overline{\eta}/\overline{\eta}^{\star})$ at its operating power (with $\overline{\eta}^{\star}$ varying from generation to generation) and then average it for each generation of power plants (shown as colored bars) with the error bars given by their respective intra-generation standard deviations. Using this multiobjective optimization criterion, we find a systematic, statistically-significant improvement in the thermal performance of the pants across successive generations, with efficiencies improving from $\overline{\eta} \simeq 0.5 \overline{\eta}^\star$ in the first generation to values close to $\overline{\eta} \simeq 0.75 \overline{\eta}^\star$ in the fourth generation.}

{\it Conclusions.---}We introduced and solved analytically a multi-objective optimization framework to characterize the fundamental performance limits of thermal machines operating in the linear response regime.
A central outcome of our analysis is the identification of universal Pareto fronts {  that set upper bounds for the performance and are achievable for endoreversible machines (i.e. under tight coupling conditions).} The geometry of these optimal fronts is independent of microscopic details and depends only on the relative weighting of the optimized objectives. {  These optimal fronts provide a unifying framework to derive universal linear-response and tradeoff relations, including specific examples of thermodynamic uncertainty relations, which are obtained as two-dimensional or three-dimensional projections of the full solution of the optimization problem.} 
We have put to the test our theoretical predictions with existing experimental data across a wide range of physical scales, from microscopic and mesoscopic engines to macroscopic heat engines and power plants. The observed performance of these real-world thermal machines falls with striking accuracy within the predicted tradeoff limits{,   even if the machines operate out of the linear response regime. This suggests that the results derived here under linear response conditions, might be extended beyond this regime, similarly to the thermodynamic uncertainty relations given by Eq.~\eqref{eq:7}, which apply arbitrary far from equilibrium.}

\bibliography{reference} 

%%%%%%%%%%%%%%%% ACKNOWLEDGEMENTS %%%%%%%%%%%%%%

\section*{Acknowledgments}
We thank John Bechhoefer for fruitful discussions on optimization problems and for granting permission to reproduce the experimental data from Ref.~\cite{Proesmans16}. We also thank Artur Widera, Roberto M. Serra, and Hui Dong for granting permission to reproduce the experimental data from Refs.~\cite{Bouton21,Peterson19,Zhai23,Cui25} in Figs.~\ref{fig: experiemntal_data} and \ref{fig: experiemntal_data_2}. We also thank Luis F. Seoane for interesting comments on Pareto fronts. We wish to acknowledge support from the CoQuSy project PID2022-140506NB-C21 and the Mar\'ia de Maeztu project CEX2021-001164-M for Units of Excellence, funded by MICIU/AEI/10.13039/501100011033/FEDER, UE. GM acknowledges the Ram\'on y Cajal program RYC2021-031121-I funded by MICIU/AEI/10.13039/501100011033 and European Union NextGenerationEU/PRTR. JAAM acknowledges Conselleria d'Educaci\'o, Universitat i Recerca of the Balearic Islands (Grant FPI$\_058\_2022$). 

%%%%%%%%%%%%%%%% SUPPLEMENT LIST %%%%%%%%%%%%%%%

%%%%%%%%%%%%%%%% END OF MAIN TEXT %%%%%%%%%%%%%%%
\clearpage
\newpage

%%%%%%%%%%%%%%%% START OF SUPPLEMENT %%%%%%%%%%%%%%%

\renewcommand{\thefigure}{S\arabic{figure}}
\renewcommand{\thetable}{S\arabic{table}}
\renewcommand{\theequation}{S\arabic{equation}}
\renewcommand{\thepage}{S\arabic{page}}
\setcounter{figure}{0}
 \setcounter{table}{0}
\setcounter{equation}{0}
\setcounter{page}{1} 

%%%%%%%%%%%%%%%% SUPPLEMENT TITLE PAGE %%%%%%%%%%%%%%%
%\begin{center}
%\appendix

\onecolumngrid
\newpage

\begin{center}
\section*{Supplemental Material for ``Universal Pareto fronts and trade-off relations in multi-objective optimization of thermal machines"}

José A. Almanza-Marrero$^{\ast}$,
Édgar Roldán,
Gonzalo Manzano\\ 
\small$^\ast$ joseantonio@ifisc.uib-csic.es\\
\end{center}

\subsection{1.Details of the optimization process for endoreversible thermal machines}

Herein, we provide the mathematical derivation for the solution of the multi-objective optimization problem.  Our starting point is Eq. (6) in the main text, copied here for convenience
\begin{equation}
\Omega(\alpha,\beta,\gamma) = \alpha~\overline{P} + \beta~\overline{\eta} - \gamma~\overline{\Sigma} - (1-\alpha-\beta-\gamma)\overline{\sigma^2_{P}}.\label{eq:omegaSM}
\end{equation}
 Because the thermodynamic variables entering the optimization procedure are normalized (and the efficiency is normalized by definition), our goal is to optimize the values of the power $P$, entropy production $\Sigma$, and power variance $\sigma_{P}^{2}$ . We compute the {\textit{ constrained}} maximum values 
 with respect to $L_{11}$ and $A_{1}$ while maintaining $L_{22}$ and $A_{2}$ fixed. We find that the constrained maximum values for the mean power, dissipation, and power fluctuations (efficiency is normalized by definition) are   respectively given by
\begin{equation}
\begin{split}
    &\max(P) = \frac{1}{4}T A_2^2 L_{22} ~~;~~ \max(\Sigma) = A_2^2 L_{22} ~~;~~ \max(\sigma_{P}^{2}) = 2 T^{2} A_2^2 L_{22}. \label{eq:s2}
\end{split}
\end{equation}

Equation~\eqref{eq:s2} leads to the following values of the thermodynamic normalized by their constrained maximum values
\begin{equation}
\begin{split}
    &\overline{P} = \frac{4 A_1 \left(A_2 \sqrt{L_{11} L_{22}}-A_1 L_{11}\right)}{A_2^2 L_{22}}, ~~~\overline{\eta} = \sqrt{\frac{L_{11}}{L_{22}}}\frac{A_1}{A_2}, ~~~ \overline{\Sigma} = \frac{\left(A_1 \sqrt{L_{11}}-A_2 \sqrt{L_{22}}\right){}^2}{A_2^2 L_{22}},~~~ \overline{\sigma^2_{P}} = \frac{A_1^2 L_{11}}{A_2^2 L_{22}}. \label{eq:s3}
\end{split}
\end{equation}

The optimization function given by Eq.~\eqref{eq:omegaSM} reads, after substitution of Eq.~\eqref{eq:s3},
\begin{equation}
   \Omega(\boldsymbol{\lambda}) =  \frac{A_2 A_1 \sqrt{L_{11} L_{22}} (4 \alpha +\beta +2 \gamma )+A_1^2 L_{11} (-3 \alpha +\beta -1)-A_2^2 \gamma  L_{22}}{A_2^2 L_{22}}. 
\end{equation}
{  At fixed $A_2$ and $L_{22}$ the objective only depends on $A_1$ and $L_{11}$, whose physical operating interval is $0\leq A_1\sqrt{L_{11}}\leq A_2\sqrt{L_{22}}$. The stationary conditions are
\begin{equation}
\nabla_{A_1,L_{11}} \Omega (\boldsymbol{\lambda}) =\left(\dfrac{\partial \Omega (\boldsymbol{\lambda})}{\partial A_1}, \dfrac{\partial\Omega (\boldsymbol{\lambda})}{\partial L_{11}}\right)= \boldsymbol{0},
\end{equation}
these conditions gives the solution $A_1 = \frac{A_2 \sqrt{\frac{L_{22}}{L_{11}}} (4 \alpha +\beta +2 \gamma )}{6 \alpha -2 \beta +2}$. This solution reaches the reversible boundary at $\beta_{\rm r} \equiv \dfrac{2(\alpha -\gamma +1)}{3}$, for larger admissible values of $\beta$, the constrained maximum remains at the boundary. Consequently the constrained solution for the relation between the optimal load $A_1$ and $L_{11}$ parameter reads:   
\begin{equation}
    A_1^{\star} = A_2\sqrt{\dfrac{L_{22}}{L_{11}^\star}} \begin{cases}
        \dfrac{4\alpha + \beta + 2\gamma}{6\alpha-2\beta +2}, \qquad &\beta< \beta_{\rm r},\\
        1,\qquad\qquad&\beta\geq \beta_{\rm r}. 
    \end{cases}
\end{equation}

Substituting the value of the stationary branch ($\beta< \beta_{\rm r}$) into Eq.~\eqref{eq:s3} gives the optimal values of the objectives as a function of the weights:
\begin{equation}\label{eq: first solution}
    \begin{split}
    &\overline{P} = \frac{(2 \alpha -3 \beta -2 \gamma +2) (4 \alpha +\beta +2 \gamma )}{(-3 \alpha +\beta -1)^2},  ~~~ \overline{\eta} = \frac{4 \alpha +\beta +2 \gamma }{6 \alpha -2 \beta +2}, ~~~ \overline{\Sigma} = \frac{(2 \alpha -3 \beta -2 \gamma +2)^2}{4 (-3 \alpha +\beta -1)^2}, ~~~ \overline{\sigma_{P}^{2}} = \frac{(4 \alpha +\beta +2 \gamma )^2}{(6 \alpha -2 \beta +2)^2}, 
\end{split}
\end{equation}
while for every admissible $\beta\geq \beta_{\rm r}$, the optimal values are instead $(\overline{P}^\star,\overline{\eta}^\star,\overline{\Sigma}^\star,\overline{\sigma_P^2}^\star) = (0,1,0,1)$. The two branches meet continuously at $\beta=\beta_{\rm r}$ and yield to Eqs. (\ref{eq:9})--(\ref{eq:12}) of the main text. 

As consistency checks, the piecewise expressions recover the optima of all four single-objective problems, obtained when all original weights vanish except the one associated with the selected objective. They reproduce maximum power at \(\overline{\eta}=1/2\), maximum efficiency and minimum dissipation at the reversible limit \(\overline{\eta}=1\), and minimum power fluctuations at \(\overline{\eta}=0\). Furthermore, the nondominated branches of the two- and three-objective projections reproduce the Pareto fronts obtained by directly solving the corresponding reduced optimization problems. These checks establish consistency with both the single-objective limits and the reduced-objective tradeoffs.}

\subsection{2.Bounding the performance of non-endoreversible thermal machines}

Here we provide explicit mathematical derivations  of the results presented in Sec. 3.2 of the Main Text. We first introduce a set of dimensionless parameters that simplify the forthcoming  formulae and derivations. First, we define the dimensionless  parameter
\begin{equation}
x \equiv -\frac{A_1}{A_2}\sqrt{\frac{L_{11}}{L_{22}}} \ge 0,
\label{eq:S3}
\end{equation}
and  the dimensionless coupling coefficients
\begin{equation}
q_1 \equiv \frac{L_{12}}{\sqrt{L_{11}L_{22}}},
\qquad
q_2 \equiv \frac{L_{21}}{\sqrt{L_{11}L_{22}}}.
\label{eq:S4}
\end{equation}
The power defined in Eq.~(1) of the Main Text [$P = -T~ A_1 J_1$] may be written, using Eqs.~\eqref{eq:S3} and \eqref{eq:S4}, as a function of $q_1$ as
\begin{equation}
P(q_1)
= T L_{22} A_2^2 x(q_1 - x).
\label{eq:S5}
\end{equation}

As discussed in the main text, the second law of thermodynamics requires the symmetric part of the Onsager matrix $L_{ij}$ to be positive semi-definite, which implies the constraint
\begin{equation}
(q_1 + q_2)^2 \le 4.
\label{eq:S6}
\end{equation}
In the presence of microscopic reversibility—e.g., in the absence of magnetic fields or non-conservative forces—Onsager reciprocity holds, yielding $L_{12} = L_{21}$ which implies
\begin{equation}
 q_1 = q_2\equiv q.
\label{eq:S7}
\end{equation}
Moreover, positivity of entropy production further implies
\begin{equation}
|L_{12}| \le \sqrt{L_{11}L_{22}}.
\label{eq:S8}
\end{equation}
Consequently, for any non-endoreversible thermal machine obeying Onsager reciprocity, $0 \le q \le 1$, while the endoreversible limit corresponds to perfect coupling, $q = 1$. In this case, the associated output power becomes
\begin{equation}
P_{\mathrm{endo}}(q=1)
= T L_{22} A_2^2 x(1 - x).
\label{eq:S11}
\end{equation}
Subtracting Eqs.~\eqref{eq:S5} and \eqref{eq:S11} gives
\begin{equation}
P_{\mathrm{endo}}(q=1) - P(q)
= T L_{22} A_2^2 x(1 - q).
\label{eq:S12}
\end{equation}
Since by definition $1 - q \ge 0$, it follows that, for all $x \ge 0$, we get
\begin{equation}
P_{\mathrm{endo}}(q=1) \ge P(q \le 1),
\label{eq:S13}
\end{equation}
with equality only for $q = 1$ (endoreversible operation) or $x = 0$ (zero power).

We now discuss the efficiency, defined in Eq.~(3) of the main text [$\overline{\eta} = -A_1 J_1/A_2 J_2$]. Expressing the fluxes in terms of the dimensionless variables introduced above, the entropy-producing flux reads
\begin{equation}
J_2 = L_{22} A_2 (1 - q_2 x),
\label{eq:S16}
\end{equation}
where $1 - q_2 x > 0$ and the inequality ensures the correct operation regime. The work-producing flux is
\begin{equation}
J_1 = \sqrt{L_{11}L_{22}} A_2 (q_1 - x).
\label{eq:S17}
\end{equation}
Using Eqs.~(3), \eqref{eq:S16}, and \eqref{eq:S17}, the efficiency reads
\begin{equation}
\eta(q_1,q_2) = 
\frac{x(q_1 - x)}{1 - q_2 x},
\qquad
x \ge 0,\quad 1 - q_2 x > 0.
\label{eq:S18}
\end{equation}
The condition $1 - q_2 x > 0$ imposes $x < 1/q_2$; under Onsager reciprocity this reduces to the same operational range $x < 1$ as for the endoreversible machine. Evaluating Eq.~\eqref{eq:S18} at $q_1 = q_2 = 1$ yields the endoreversible efficiency
\begin{equation}
\eta_{\mathrm{endo}}(q=1) = 
\frac{x(1 - x)}{1 - x}=
x,
\qquad
0 \le x < 1.
\label{eq:S23}
\end{equation}
From Eq.~\eqref{eq:S18}, for $x > 0$ the inequality $\eta \le x$ is equivalent to
\begin{equation}\label{eq: S21}
 q_1\le 1+(1-q_2)x.
\end{equation}
Under Onsager reciprocity this condition becomes $q \le 1 + (1 - q)x$, which is always satisfied in the operational regime since $q \le 1$. Equality is attained only in the endoreversible case $q = 1$. Hence, the endoreversible thermal machine achieves, at every operating point $x$, the maximum efficiency allowed by linear irreversible thermodynamics under Onsager reciprocal relations.

We next show that the endoreversible thermal machine also provides a lower bound for the entropy production rate. Starting from the definition of the entropy production rate given in Eq.~(2) of the main text [${\Sigma} = A_1 J_1 + A_2 J_2$] and substituting $A_1 = -x A_2 \sqrt{L_{22}/L_{11}}$, we obtain
\begin{equation}
\Sigma(q_1,q_2) = 
L_{22} A_2^2[1 - (q_1 + q_2)x + x^2].
\label{eq:S27}
\end{equation}
The endoreversible operation is again characterized by perfect coupling and Onsager reciprocity, i.e., $q = q_1 = q_2 = 1$. In this case,
\begin{equation}
\Sigma_{\mathrm{endo}}(q=1)=
L_{22} A_2^2 (1 - x)^2,
\label{eq:S30}
\end{equation}
which vanishes only in the reversible limit $x \to 1$. Subtracting Eq.~\eqref{eq:S30} from Eq.~\eqref{eq:S27} yields
\begin{align}
\Sigma(q_1,q_2) - \Sigma_{\mathrm{endo}}(q=1)
&=
L_{22} A_2^2 x(2 - q_1 - q_2).
\label{eq:S31}
\end{align}
Under Onsager reciprocity this simplifies to
\begin{equation}
\Sigma(q \le 1) - \Sigma_{\mathrm{endo}}(q=1) = 
2 L_{22} A_2^2 x(1 - q),
\label{eq:S32}
\end{equation}
since $0 \le q \le 1$ and $0 \le x \le 1$ in the operational domain, we conclude that
\begin{equation}
\Sigma(q \le 1) \ge \Sigma_{\mathrm{endo}}(q=1),
\label{eq:S33}
\end{equation}
for all $x \in [0,1]$ with equality only for $q = 1$ (endoreversible operation) or $x = 0$. Thus, at every operating point $x$, the endoreversible thermal machine yields the minimal entropy production rate among all linear-response machines obeying Onsager reciprocal relations.

Finally, taking into account that fluctuations in the output power do not depend on the off-diagonal Onsager coefficients, together with our demonstration that endoreversible operation always bounds the performance of non-endoreversible machines satisfying Onsager symmetry, we conclude that the optimal Pareto fronts derived for endoreversible machines represent the ultimate performance limits for non-endoreversible reciprocal machines. We can relax the condition of Onsager symmetry and determine for which values of \(q_1\) and \(q_2\) the endoreversible operation still bounds the non-endoreversible one. Starting from the condition on the power, we find that \(q_1 \le 1\). Using this condition together with the one in \eqref{eq: S21}, we also obtain \(q_2 \le 1\). The two conditions obtained so far, when considered together, also satisfy \eqref{eq:S31}.

\subsection{3. Pareto fronts under resource optimization}

In the main text, we determine the optimal trade-offs for machines operating with fixed thermodynamic resources, specified by $A_2$ and $L_{22}$. Here, we consider the distinct problem in which these resources are also allowed to vary. The weighted-sum method employed in Eq.~(6) of the main text does not yield a finite optimum in this setting: for any nonzero power weight $\alpha$, varying the resources allows configurations with arbitrarily large power, leading to $\Omega\to\infty$. Nevertheless, the performance bound established in Eq.~(7) allows us to demonstrate that endoreversible machines remain Pareto optimal, as shown below.

An endoreversible configuration $\mathbb C'=(A_1',A_2',L_{11}',L_{22}')$ dominates another endoreversible configuration $\mathbb C=(A_1,A_2,L_{11},L_{22})$ if
\begin{equation}
    P' \geq P, \qquad
    \eta' \geq \eta, \qquad
    \Sigma' \leq \Sigma, \qquad
    \sigma_{P'}^2 \leq \sigma_P^2,
\end{equation}
with at least one strict inequality. Here, primed and unprimed quantities are evaluated at $\mathbb C'$ and $\mathbb C$, respectively. Every endoreversible machine satisfies $2P^2=\Sigma\sigma_P^2$. Assuming strictly positive objective values, the dominance conditions therefore imply
\begin{equation}
    2P^2
    \leq 2(P')^2
    = \Sigma'\sigma_{P'}^2
    \leq \Sigma\sigma_P^2
    = 2P^2.
\end{equation}
Because the chain begins and ends with the same value, every inequality must be an equality. Consequently,
\begin{equation}
    P'=P,
    \qquad
    \Sigma'\sigma_{P'}^2=\Sigma\sigma_P^2.
\end{equation}
The equality of these products, together with $\Sigma'\leq\Sigma$ and $\sigma_{P'}^2\leq\sigma_P^2$, requires both factors to remain unchanged. Indeed, since both factors are strictly positive, decreasing either one would require increasing the other to preserve the product, violating the dominance conditions. Thus,
\begin{equation}
    \Sigma'=\Sigma,
    \qquad
    \sigma_{P'}^2=\sigma_P^2.
\end{equation}
At a fixed reference temperature $T$, the efficiency defined in the main text satisfies
\begin{equation}
    \eta=\frac{P}{P+T\Sigma}.
\end{equation}
Hence, $P'=P$ and $\Sigma'=\Sigma$ also imply $\eta'=\eta$. All four objective values must therefore coincide, contradicting the requirement of at least one strict improvement. Endoreversible configurations are consequently mutually nondominating.

Having established that endoreversible configurations are mutually nondominating, we use the analytical result in Eq.~(7) of the main text to show that these machines also determine the limits of optimal performance for non-endoreversible machines. These limits are expressed by
\begin{equation}
    \Sigma\,\frac{\sigma_P^2}{P^2}\geq 2,
\end{equation}
which coincides with the bound obtained for fixed thermodynamic resources.

This result, however, does not provide a full parametric solution to the optimization problem when the resources are allowed to vary. In particular, it does not express the optimal objective values in terms of the weights assigned to each objective, nor does it determine the corresponding optimal configuration parameters.

\end{document}